\documentclass[10pt,conference]{IEEEtran}

\usepackage{enumitem}
\usepackage{footnote}
\usepackage{amsmath}
\usepackage{array}
\usepackage{graphicx}
\usepackage[dvipsnames]{xcolor}
\usepackage{lscape}
\usepackage{subcaption}
\usepackage{soul}
\usepackage{tablefootnote}
\usepackage{hyperref}

\usepackage[switch]{lineno}

\def\BibTeX{{\rm B\kern-.05em{\sc i\kern-.025em b}\kern-.08emT\kern-.1667em\lower.7ex\hbox{E}\kern-.125emX}}

\newcommand{\edit}[1]{\textcolor{black}{#1 }}

\graphicspath{ {figures/} }

\usepackage[labelfont=bf]{caption}
\usepackage{multirow}

\begin{document}

\title{Search4Code: Code Search Intent Classification Using Weak Supervision}

\author{
\IEEEauthorblockN{Nikitha Rao\IEEEauthorrefmark{1}, Chetan Bansal\IEEEauthorrefmark{1}, Joe Guan\IEEEauthorrefmark{2}}
\IEEEauthorblockA{\IEEEauthorrefmark{1}\textit{Microsoft Research, }\IEEEauthorrefmark{2}\textit{Microsoft}}
\IEEEauthorblockA{\{t-nirao, chetanb, zhgua\}@microsoft.com}
}

\maketitle
\begin{abstract}
Developers use search for various tasks such as finding code, documentation, debugging information, etc. In particular, web search is heavily used by developers for finding code examples and snippets during the coding process. Recently, natural language based code search has been an active area of research. However, the lack of real-world large-scale datasets is a significant bottleneck. In this work, we propose a weak supervision based approach for detecting code search intent in search queries for C\# and Java programming languages. We evaluate the approach against several baselines on a real-world dataset comprised of over 1 million queries mined from Bing web search engine and show that the CNN based model can achieve an accuracy of 77\% and 76\% for C\# and Java respectively. Furthermore, we are also releasing Search4Code, the first large-scale real-world dataset of code search queries mined from Bing web search engine. We hope that the dataset will aid future research on code search.
\end{abstract}

\section{Introduction}

Searching for code is a common task that developers perform on a regular basis. There are many sources that developers use to search for code: web search engines, code repositories, documentation, online forums, etc. Code searches typically contain a query composed of natural language and expect a code snippet result. Natural language based code search has been looked at by different approaches such as traditional information retrieval techniques \cite{ir1,ir2,ir3}, deep learning \cite{deepcs}, and hybrid approaches \cite{ncs} that combine various methodologies. One commonality that exists is the requirement of a sufficiently large dataset composed of code and the corresponding natural language labels. Traditionally, researchers have used different methods to gather data, including using the associated docstring of the code snippet and the question title from coding related forums (e.g. StackOverflow). However, these natural language labels do not accurately represent how developers perform searches for code in a typical search engine. While there exist some datasets that include human-annotated labels for code \cite{codesearchnet}, these are limited in size and quantity. 

We present a dataset compiled from query logs comprised of millions of queries from Bing web search engine. This dataset contains aggregated queries, which have been anonymized, and classified as either having a code search intent or not for C\# and Java programming languages. The dataset also contains the most frequently clicked URLs and a popularity metric denoting the query frequency. To create a large-scale dataset of code search queries, it is crucial to automatically detect code search intent in search queries. Previous research in the area of search query classification \cite{sesearchintent,evaluatingsearch} has focused primarily on the classification of web queries in categories such as Debug, API, and HowTo using heuristics and rule-based methods which tend to overfit. 

In this paper, we introduce a novel weak supervision based model to classify code search intent in search queries. We define a query as having code search intent if it can be sufficiently answered with a snippet of code. 
To the best of our knowledge, this is the first usage of weak supervision in the software engineering domain. In summary, our main contributions are:
\vspace{-0.7mm}
\begin{itemize}[leftmargin=4mm]
    \item A novel weak supervision based model to detect code search intent in queries.
    \item A large-scale dataset of queries\footnote{\url{https://github.com/microsoft/Search4Code/}}, mined from Bing web search engine, that can be used for code search research.
\end{itemize}

\section{Background and Motivation}
\label{sec:background}
Our work builds on recent advances in the areas of code search, search query intent classification, and weak supervision. In this section, we provide a brief background of the same. \\

\noindent\textbf{Code Search:} Code search is a sub-field in \edit{information retrieval} that focuses on finding relevant code snippets given a natural language query. Code search is an integral part of the software development process \cite{codesearch1,codesearch2,codesearch3} as developers often search for code using search engines, documentation, and online forums. However, a significant bottleneck in this area is the lack of datasets for building and experimenting with new techniques. The most recent work in curating a dataset contains $99$ human-annotated queries across multiple languages \cite{codesearchnet} and $287$ question-answer pairs extracted from StackOverflow \cite{ncsevaluation}. We aim to contribute a new method to generate a code search dataset by mining query logs from Bing web search engine. Additionally, we open-source this dataset to aid future research on code search.\\

\begin{figure*}[t!]
\vspace{-8mm}
\begin{center}
\includegraphics[width=0.75\linewidth]{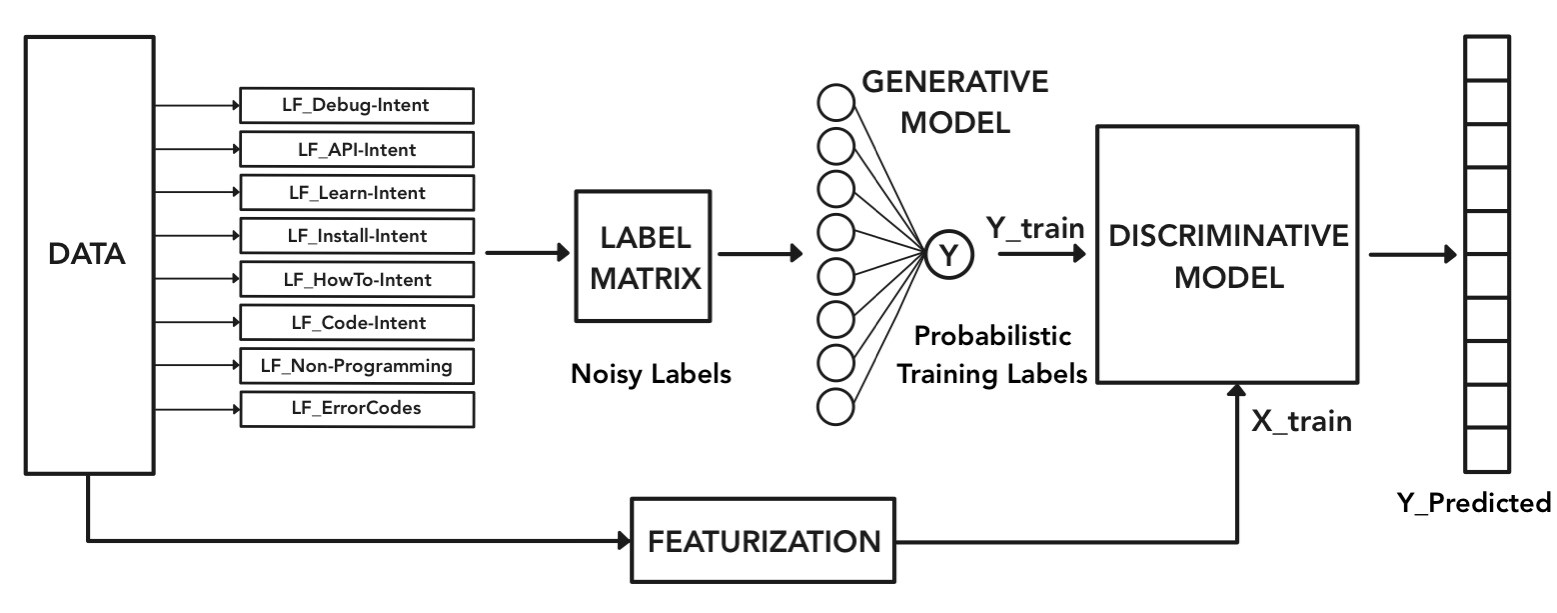}
\vspace{-3mm}
\caption{Overview of the pipeline.}
\label{fig:pipeline}
\end{center}
\vspace{-9mm}
\end{figure*}

\noindent\textbf{Intent Classification:} Applications of intent classification in web search include several domains like healthcare~\cite{health}, security~\cite{bansalsigir} and e-commerce \cite{ahuja2020language, rao2020product}. Wang et al. have leveraged intent understanding for improving effort estimation in code reviews \cite{wang2019leveraging, wang2020large}. Recently, software engineering related search queries have been analyzed and classified into different categories by using distant supervision \cite{sesearchintent} and token-level intent aggregation \cite{evaluatingsearch} to better understand developer behaviour. Our goal is to further improve upon these methods by introducing a weak supervision based approach for code search intent classification. \\

\noindent\textbf{Weak Supervision:} One of the primary challenges in supervised learning is to obtain large-scale labelled data. As mentioned above, this obstacle exists in the code search space as well. Weak supervision \cite{weaksupervision, weaksupervision2} leverages `weak' or `noisy' learning functions to automatically assign labels to a large amount of unlabeled data.

Formally speaking, given a set of unlabeled data points, $X$, the objective of weak supervision is to estimate the ground truth label by using a set of $n$ learning functions. Each learning function has a probability to abstain and a probability to correctly label a data point as positive or negative.
The learning functions are applied over $m$ unlabeled data points to create a matrix of label outputs, $\Lambda$. The generative model then takes $\Lambda$ as input and returns the probability scores for each class based on the agreements and disagreements between the learning functions. 
The predicted label distribution output can then be used as probabilistic training labels by a discriminative classifier for a downstream classification task. We use weak supervision to generate the train labels for the code search intent classification task.

\section{Approach}

In this section, we elaborate on our approach for code intent classification. First, we build the generative model using weak supervision to get the labels for the training data using snorkel, a weak supervision framework by Stanford~\cite{snorkel}. We then use this data to train discriminative models to classify queries as having code search intent or not. Figure~\ref{fig:pipeline} provides an overview of the entire pipeline.

\subsection{Generative Model Pipeline}
\label{sec:generative-model-pipeline}

\noindent \textbf{Data Collection:} We randomly sample $1$ million search queries each for C\# and Java, collected from $1^{st}$ September, $2019$ to $31^{st}$ August, $2020$ from Bing web search engine. We identify queries related to each programming language by doing a simple keyword-based pattern matching (`c\#', `c sharp' and `csharp' for C\# and `java' for Java)~\cite{hassan2020empirical}. We apply additional filters to ensure that all the queries are in English locale from the USA region and we eliminate any traffic from bots and other services. Additionally, we exclude queries that have multiple programming languages in them such as `c\# vs java', `how hard is c\# compared to java or c++?', `java to c\# converter' and so on to better isolate queries to an individual programming language.\\

\noindent \textbf{Learning Functions (LFs):} As discussed in Section~\ref{sec:background}, we use several `weak' or `noisy' learning functions, described in Table~\ref{table:learning-functions}, that are combined in a weighted manner by the generative model. Weak supervision sources generally include external knowledge bases, patterns, dictionaries and even domain-specific heuristics. In the context of code search intent classification, we leverage the software engineering sub-intent classifiers (such as Debug, HowTo, etc.) proposed by Rao et al. \cite{sesearchintent}. We also introduce learning functions to identify patterns that indicate code examples, error codes and exceptions. Each learning function acts as a binary classifier that identifies either code search or not code search intent and abstains otherwise. We use the label $1$ for code search intent, $0$ for not code search intent and $-1$ for abstain. 
The label for each learning function is chosen after manually analyzing a sample of queries. Table~\ref{table:learning-functions} provides the target label and description of heuristics used for each of the learning functions used along with a few example queries. \\

\begin{table*}
\vspace{-8mm}
\small
  \begin{center}
    \begin{tabular}{|l|c|p{6.5cm}|p{5.5cm}|}
    \hline
    \textbf{Learning Function} & \textbf{Label} & \textbf{Description} & \textbf{Examples} \\
    \hline
    API Intent & 1 & Keywords like `api',`function',`method',`call', etc. that indicate a specific API usage. & `c\# example of restful post api call form url encode', `java immutablelist api'.\\
    \hline
    Debug Intent & 0 & Keywords like `error',`exception',`fail',`not working', `debug', etc. that indicate an error or issue. &  `500 internal server error in web api c\#', `java createnewfile not working'.\\
    \hline
    HowTo Intent & 1 & Keyword `how' is present to indicate the need to accomplish a specific task.  &  `c\# asp.net how to implement click event for textbox', `how to do quicksort in java'.\\
    \hline
    Learn Intent & 0 & Keywords like `tutorial',`what',`why',`difference', `versus', etc. that indicate learning new topics.  & `block body vs lambda method c\#', `what is the order of precedence for java math'.  \\
    \hline
    Install Intent & 0 & Keywords like `install', `download', `update' etc. that indicate installing software packages. & `c\# .net install .msi remotely', `download selenium web driver jars for java'  \\
    \hline
    Code Search Intent & 1 & Keywords like `example',`sample code',`snippet', `implementation', etc. that indicate code search. & `proxysocket c\# code sample', `java void method no parameters example' \\
    \hline
    Non-Programming & 0 & Keywords like `interview', `jobs', etc. that indicate non-programming related queries. & `c\# array questions for interviews', `part time java coding jobs'\\
    \hline
    Error Codes & 0 & Regex based patterns to find C\# error codes or Java exceptions & `cs7038 wcf c\# failed to emit module', `java.io.eofexception: postman'.\\
    \hline
    \end{tabular}
    \caption{Overview of the learning functions.}
    \label{table:learning-functions}
  \end{center}
\vspace{-8mm}
\end{table*}
\noindent \textbf{Generative Model:} We apply all the individual learning functions to the data and construct a label matrix that is then fed to the generative model. The generative model then uses a weighted average of all learning functions outputs, based on the agreements and disagreements between the learning functions, to return the probability scores for each class. Each datapoint is then assigned a label based on the class having the higher probability score. 

\subsection{Discriminative Model Pipeline}
\label{sec:discriminative-model-pipeline}
\noindent \textbf{Data:} We use the output of the generative model as the train labels (Y\_train) for the data we collected earlier. We then preprocess and featurize the data before passing it to the discriminative model. \\

\noindent \textbf{Preprocessing and Featurization:} We first tokenize the queries based on non-alphanumeric characters and remove all stopwords. We then transform the query text into its vector representation using Word2Vec~\cite{Mikolov:2013:DRW:2999792.2999959} to capture any semantic similarities. We retrain the Word2Vec model from scratch on our query data since the pre-trained Word2Vec models don't generalize well to queries related to programming languages. We compute the word embeddings for each token in a query using the trained Word2Vec model and compute query embedding as the average of all token embeddings. This forms the training data (X\_train) for the discriminative models.\\

\noindent \textbf{Discriminative Model:} Using the generated training labels (Y\_train) along with the featurized train data (X\_train) data, we train several supervised machine learning and deep learning models to tackle the problem of code search intent detection in search queries. We further elaborate on the various discriminative models used in Section~\ref{sec:experimental-setup}.

\section{Experiments and Results}
In this section, we first describe the experimental setup. We then present the evaluation for the generative model that is used to derive train data labels. Lastly, we evaluate the efficacy of various discriminative models for code search intent classification in search queries. We evaluate the performance of each model against the overall test accuracy along with the precision, recall and F1 scores for both classes. Note that we train and evaluate the models for C\# and Java separately.

\subsection{Experimental Setup}
\label{sec:experimental-setup}
\noindent \textbf{Dataset:} The featurized data described in Section~\ref{sec:discriminative-model-pipeline} along with the generated train labels from Section~\ref{sec:generative-model-pipeline} is used as the training data for the various discriminative models.

For the test data, we uniformly sample a set of $200$ queries based on query length for both C\# and Java. Three annotators then manually label the data independently. We compute the inter-rater agreement score to be $0.75$ using Fleiss' Kappa~\cite{fleiss1971measuring}, which translates to substantial agreement. The final label is obtained by taking a majority vote. We find the distribution of queries with code search intent in the manually labelled test data to be $62.0$\% for C\# and $34.5$\% for Java.\\

\noindent \textbf{Discriminative Models:} We compare the performance of various machine learning and deep learning models to find the best performing code search intent classification model. In particular, we look at the following discriminative models
\begin{itemize}[leftmargin=4mm]
    \item First, we look at non-deep learning models like Logistic Regression and Random Forest. We use the default version of the models from scikit-learn to implement them.
    \item For the deep learning models, we look at Bidirectional LSTM (BiLSTM) with attention and CNN. The BiLSTM is implemented by adding the bidirectional layer on top of the LSTM layer~\cite{biLSTM}. For the CNN, we use convolution layers with ReLu activation followed by maxpool layers and a dense output layer with sigmoid activation~\cite{CNN-blog, le2017convolutional}. We implement the models using keras with tensorflow backend.
\end{itemize}

\subsection{Generative Model Evaluation}
To evaluate the performance of the generative model for generating the train data labels, we compare the performance of the model with a majority vote model on the test data. The majority vote model assigns the label for each query based on the majority vote of all eight learning functions and ties are settled by assigning a random label. Table~\ref{table:generative-model-evaluation} summarizes the evaluation scores for the two models. We find that the generative model outperforms the majority vote model across all metrics with an overall test accuracy of $73\%$ and $72\%$ for C\# and Java respectively.

\begin{table*}
\vspace{-4mm}
\small
\begin{center}
    \begin{tabular}{|c|c|c|c|c|c|c|c|c|}
        \hline
         \multirow{2}{*}{\parbox{1.9cm}{\textbf{Programming\\Language}}} & \multirow{2}{*}{\textbf{Model}} & \multirow{2}{*}{\textbf{Accuracy}}& \multicolumn{3}{|c|}{\textbf{Code Intent}} & \multicolumn{3}{|c|}{\textbf{Not Code Intent}}\\\cline{4-9}
          &  &  & \textbf{Precision} & \textbf{Recall} & \textbf{F1 Score} & \textbf{Precision} & \textbf{Recall} & \textbf{F1 Score}\\
        \hline
        \multirow{2}{*}{C\#} & {Majority Vote} & 66 & 73 & 72 & 72 & 55 & 57 & 56 \\
        \cline{2-9}
         & {Generative} & \textbf{73} & \textbf{80} & \textbf{76} & \textbf{78} & \textbf{63} & \textbf{68} & \textbf{66} \\
        \hline
        \multirow{2}{*}{Java} & {Majority Vote} & 67 & 52 & 71 & 60 & 81 & 65 & 72 \\
        \cline{2-9}
        & {Generative} & \textbf{72} & \textbf{57} & \textbf{80} &\textbf{67} & \textbf{87} & \textbf{68} & \textbf{76} \\
        \hline
        \end{tabular}
        \caption{Evaluation of the generative model on the test data.}
        \label{table:generative-model-evaluation}
\vspace{-2mm}
\end{center}
\end{table*}

\begin{table*}
\vspace{-2mm}
\small
\begin{center}
    \begin{tabular}{|c|c|c|c|c|c|c|c|c|}
        \hline
         \multirow{2}{*}{\parbox{1.9cm}{\textbf{Programming\\Language}}} & \multirow{2}{*}{\textbf{Model}} & \multirow{2}{*}{\textbf{Accuracy}}& \multicolumn{3}{|c|}{\textbf{Code Intent}} & \multicolumn{3}{|c|}{\textbf{Not Code Intent}}\\\cline{4-9}
          &  &  & \textbf{Precision} & \textbf{Recall} & \textbf{F1 Score} & \textbf{Precision} & \textbf{Recall} & \textbf{F1 Score}\\
        \hline
        \multirow{5}{*}{C\#} & {Logistic Regression} & 71 & 73 & 86 & 79 & 68 & 47 & 56 \\
        \cline{2-9}
         & {Random Forest} & 73 & 73 & \textbf{90} & 80 & \textbf{72} & 45 & 55 \\
        \cline{2-9}
         & {CNN} & \textbf{77} & \textbf{79} & 85 & \textbf{82} & \textbf{72} & \textbf{63} & \textbf{67} \\
        \cline{2-9}
         & {BiLSTM} & 72 & 77 & 78 & 78 & 64 & 62 & 63 \\
        \hline
        \multirow{5}{*}{Java} & {Logistic Regression} & 74 & 59 & 85 & \textbf{70} & 90 & 69 & 78 \\
        \cline{2-9}
         & {Random Forest} & 73 & 57 & \textbf{89} & \textbf{70} & \textbf{91} & 65 & 76 \\
        \cline{2-9}
         & {CNN} & \textbf{76} & \textbf{63} & 74 & 68 & 85 & \textbf{77} & \textbf{81} \\
        \cline{2-9}
         & {BiLSTM} & 73 & 59 & 76 & 66 & 85 & 72 & 78 \\
        \hline
        \end{tabular}
        \caption{Evaluation of the discriminative models on the test data.}
        \label{table:discriminative-model-evaluation}
\end{center}
\vspace{-8mm}
\end{table*}
\subsection{Discriminative Model Evaluation}
Te evaluate the efficacy of the various discriminative models for code search intent detection, we first train each model on the train data and compare the performance scores on the test data. Table~\ref{table:discriminative-model-evaluation} summarizes the performance scores of the four models. We find that the CNN model outperforms all the other models across majority of the metrics with an overall test accuracy of $77\%$ and $76\%$ for C\# and Java respectively.

\section{Code Search Query Dataset}
In this work, we have built a code search intent classification model based on weak supervision. One of the major impediments for research in this domain is the lack of publicly available large-scale datasets. On this account, we are releasing Search4Code \cite{sampleUrl}, the first large-scale real-world dataset of code search queries for C\# and Java mined from Bing web search engine. The dataset is composed of about $4,974$ C\# queries and $6,596$ Java queries. We hope that this dataset will aid future research to not just better code search intent detection but also applications like natural language based code search.

To build the dataset we first collect the anonymized query logs for one year. We then mine the code search queries by following several steps of log mining, processing and aggregation. First, we apply the same filters for locale, bots, etc. and filter out queries that are not related to C\# or Java programming languages as described in Section~\ref{sec:generative-model-pipeline}. Next, we apply a $k$-anonymity filter \cite{sweeney2002k} with a high value of $k$. This filters out queries from the dataset which were entered by less than $k$ users and could potentially contain sensitive information which was known to less than $k$ users. Finally, we apply the best performing discriminative model (i.e. CNN) to the queries to identify queries with code search intent.

We have defined the schema for the dataset in Table \ref{table:schema}. It contains not only the raw queries but also other useful attributes such as top click URLs and rank based on popularity. \edit{Here are some of the unique features of the dataset:}

\begin{itemize}
    \item \textbf{Real queries}: The queries are sampled from anonymized Bing search logs. We believe this provides a rich dataset indicative of real-world user behavior.
    \item \textbf{Click URLs}: Each query has a list of the three most frequently clicked URLs from the query logs based on user interactions.
    \item \textbf{Popularity score}: Each query is assigned a popularity rank based on the frequency of occurrence.
    \item \textbf{Large scale}: The dataset contains thousands of queries, hence, enabling large scale analysis of search for software engineering.
    \item \textbf{Other applications}: Since the dataset contains both code-search and non code-search queries, it could also be used to analyze other user intents, as described in our prior work~\cite{sesearchintent}.
\end{itemize}

\noindent\edit{\textbf{Limitations:} While we provide the top clicked URLs for each query, the code samples themselves are not provided and will have to be mined from the URLs present. Also, we predict the code search intent solely based on the search queries since we don't have access to the content from the clicked web pages. So, it is possible that not all the Clicked Urls contain code samples. Additionally, since the code search intent labels are generated by the CNN model, they will not be 100\% accurate.}

\begin{table}
\vspace{2mm}
\small
\begin{center}
\begin{tabular}[t]{| p{3.1cm} | p{4.7cm} |} 
\hline
\textbf{Attribute} & \textbf{Description}\\
\hline
Id & Identifier for the query. \\
\hline
Query & The raw query issued by the users. \\
\hline
Is Code Search Query & Whether the query has code search intent.\\
\hline
Top Clicked URLs & Top 3 document URLs (comma delimited) by click frequency. \\
\hline
Popularity Rank & Rank based on the query frequency. Most popular query is ranked 1.\\
\hline
\end{tabular}
\caption{Schema of the code search queries dataset}
\label{table:schema}
\vspace{-6mm}
\end{center}
\end{table}

\section{Conclusion And Future Work}
Search is heavily used by developers for various tasks during the software development process. Given the lack of labelled data, we use weak supervision for code search intent classification. We develop a CNN based model for code search intent classification for C\# and Java search queries mined from Bing web search engine. We also evaluate it against various baselines which demonstrates the efficacy of the weak supervision based approach. Furthermore, we are releasing the first large-scale real-world code search query dataset comprising more than $11,000$ search queries. Our code search intent model can be integrated with several applications such as IDEs, Search Engines and even developer forums like StackOverflow for improving the code search experience.

Future work on code search can leverage the dataset for building and improving natural language based code search techniques. Additionally, to the best of our knowledge, this is the first work to explore the usage of weak supervision in the software engineering domain. \edit{Weak supervision can also be leveraged in other tasks such as bug detection and program repair where a limited amount of labelled data is available.} Lastly, we plan to experiment with other advanced transformer-based neural model architectures such as BERT \cite{devlin2018bert} to improve the discriminative model performance for code search intent classification. 

\section{acknowledgements}
\label{sec:acknowledgements}
We would like to acknowledge the invaluable contributions of Mark Wilson-Thomas, Shengyu Fu, Nachi Nagappan, Tom Zimmermann and B. Ashok.

%
\bibliographystyle{IEEEtran}
\bibliography{references}

\end{document}